\documentclass[fleqn,usenatbib]{mnras}

\usepackage{newtxtext,newtxmath}

\usepackage[T1]{fontenc}
\usepackage{pdflscape}

\DeclareRobustCommand{\VAN}[3]{#2}
\let\VANthebibliography\thebibliography
\def\thebibliography{\DeclareRobustCommand{\VAN}[3]{##3}\VANthebibliography}

\usepackage{graphicx}	
\usepackage{amsmath}	

\newcommand{\alphafe}{[$\alpha$/Fe]}
\newcommand{\mgfe}{[Mg/Fe]}
\newcommand{\feh}{[Fe/H]}
\newcommand{\fehalphafe}{[Fe/H]-[$\alpha$/Fe]}
\newcommand{\mgmnalmn}{[Mg/Mn]-[Al/Fe]}
\newcommand{\msol}{$\rm M_\odot$}

\title[\alphafe{}-rise is signature of gas accretion]{The dawn is quiet here: Rise in \alphafe{} is a signature of massive gas accretion that fueled proto-Milky Way}

\author[B. Chen et al.]{
Boquan Chen,$^{1, 3}$\thanks{E-mail: boquan.chen@anu.edu.au}
Yuan-Sen Ting,$^{1, 3, 4, 5}$
Michael Hayden$^{2, 3}$
\\
$^{1}$Research School of Astronomy and Astrophysics, Australian National University, Canberra, ACT 2611, Australia\\
$^{2}$Sydney Institute for Astronomy, School of Physics, The University of Sydney, NSW 2006, Australia\\
$^{3}$ARC Centre of Excellence for All Sky Astrophysics in Three Dimensions (ASTRO-3D)\\
$^{4}$Research School of Computer Science, Australian National University, Acton ACT 2601, Australia \\
$^{5}$Department of Astronomy, The Ohio State University, Columbus, OH 43210, USA }

\date{Accepted XXX. Received YYY; in original form ZZZ}

\pubyear{2022}

\begin{document}
\label{firstpage}
\pagerange{\pageref{firstpage}--\pageref{lastpage}}
\maketitle

\begin{abstract}
The proto-Milky Way epoch forms the earliest stars in our Galaxy and sets the initial conditions for subsequent disk formation. Recent observations from APOGEE and H3 surveys showed that the \alphafe{} ratio slowly declined between \feh{} $=-3$ and $-1.3$ until it reached the lowest value ($\sim 0.25$) among the selected in situ metal-poor stars that most likely formed during the proto-Galaxy epoch. \alphafe{} rose to meet the traditional high value commonly associated with the thick disk population at \feh{} $=-1$. It was suggested that the rise in \alphafe{} could be caused by an increase in the star formation efficiency (SFE), known as the ``simmering'' phase scenario. However, gas inflow also plays a vital role in shaping the star formation history and chemical evolution of galaxies. We investigate this unexpected \alphafe{}-rise with a statistical experiment involving a galactic chemical evolution (GCE). Our model has five free parameters: the mass of the initial reservoir of the cold interstellar medium (ISM) at birth, the frequency of Type Ia supernovae (SNe Ia), the cooling timescale of the warm ISM, the SFE, and the inflow rate of fresh gas. The last two free parameters were allowed to change after \alphafe{} reached its lowest value, dividing the proto-Galaxy epoch into two phases. We find that the rise in \alphafe{} is caused by a large inflow of fresh gas and conclude that the \alphafe{}-rise is a signature of the cold mode accretion whose materials formed the prototype Milky Way preceding disk formation. Although the SFE is essential in regulating the chemical evolution, it does not necessarily increase to facilitate the \alphafe{}-rise. 
\end{abstract}




\section{Introduction}
\label{intro}

The Milky Way galaxy is a complex and dynamic system that has undergone a long and rich history of formation and evolution. One of the main goals of Galactic archaeology is to reconstruct this history by studying the properties of its stellar populations, especially the oldest and most pristine ones. Stars carry valuable information about the physical and chemical conditions of their birth environments. By measuring their photospheric elemental abundances, we can infer the nucleosynthesis processes that enriched the interstellar medium (ISM) at the time of their birth, the star formation rates (SFR), the mixing and transport mechanisms, and the merger events that shaped the Galaxy. However, it remains a challenge to identify the ancient in situ stars that formed in the pre-disk phase of our Galaxy. The elemental abundances of these stars are expected to reveal the initial conditions of our Galaxy that set the stage for disk formation. 

Stellar photospheric elemental abundances are one of the most powerful tools for Galactic archaeology, as they {reflect the gas conditions at the birth of stars and} provide direct and robust constraints on the chemical evolution of the Galaxy \citep{2002ARA&A..40..487F}. Different elements are produced by different sources, such as massive stars, Type Ia supernovae (SNe Ia), asymptotic giant branch (AGB) stars, or neutron star mergers, with different delay timescales and efficiencies \citep{kobayashi_galactic_2006, 2020ApJ...900..179K}. Certain elements, such as oxygen, neon, magnesium, silicon, sulfur, argon, calcium, and titanium, can be produced at early times in core-collapse events at relatively constant rates with iron. Stars with high \alphafe{} tend to form early when core-collapse supernovae (CCSN) efficient at producing $\alpha$-elements dominate nucleosynthesis. As time passes and Type Ia supernovae "turn on," the \alphafe{} ratio drops as \feh{} increases. The relative abundances of different elements can thus reflect the relative contributions of these separate sources as well as the time delay between their production and their incorporation into new generations of stars. 

\begin{figure*}
\centering
  \includegraphics[width=1\linewidth]{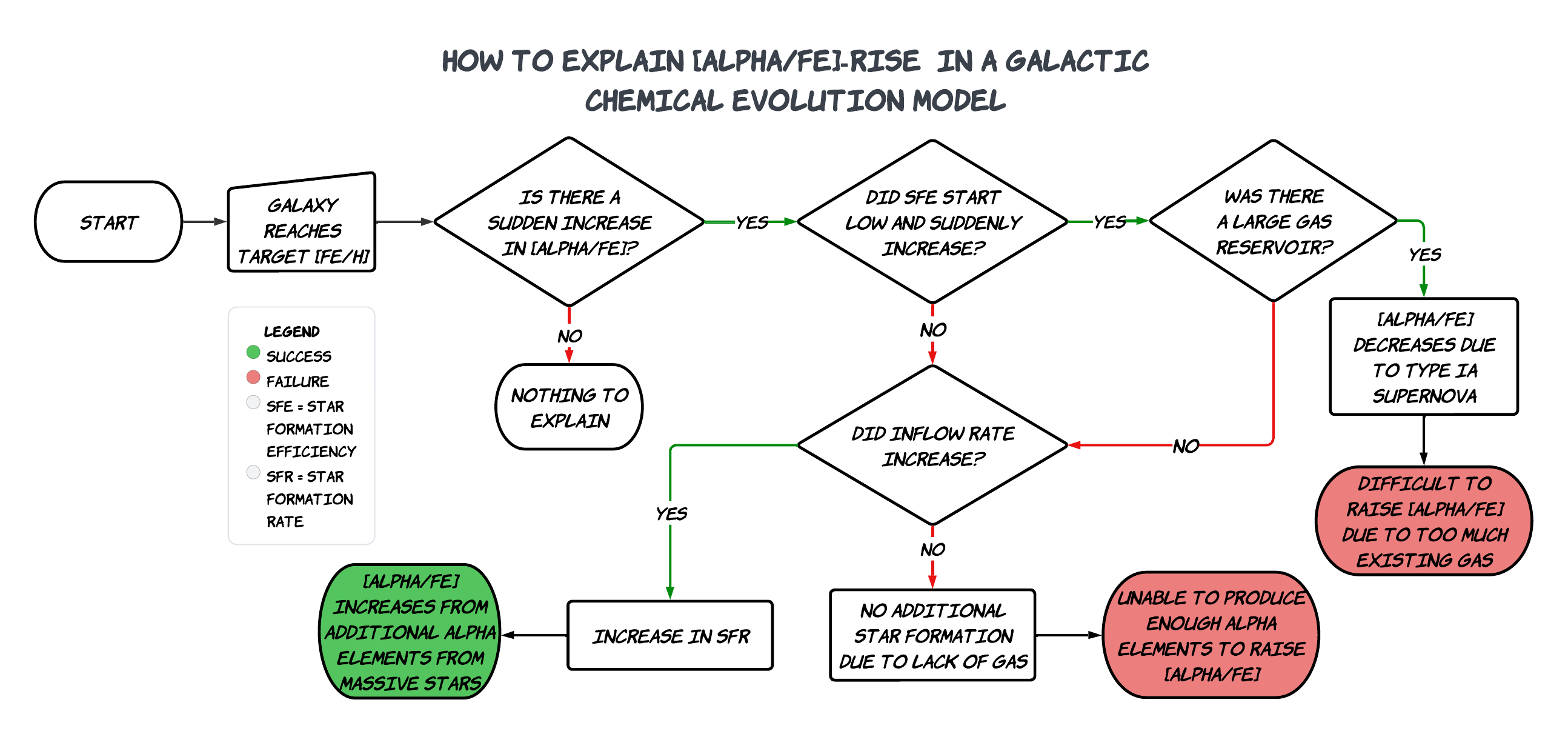}
  \caption{Flowchart illustrating scenarios of early Milky Way chemical evolution. The scenario capable of producing an \alphafe{}-rise is highlighted in green and the rest in red. In summary, the additional star formation required to raise \alphafe{} can be achieved by increasing the SFE, inflow rate, or both. However, increasing the SFE is ineffective if no gas sustains star formation. If the gas already exists as a massive gas reservoir before the parameter change, it is difficult to change the abundance in the model. It is preferable for the inflow to join the model after the parameter change. }
  \label{fig:flowchat_alpha_fe_rise}
\end{figure*}

Massive disk galaxies like the Milky Way are expected to have an ancient, metal-poor, and centrally concentrated stellar population, reflecting the star formation and enrichment in the most massive progenitor components that formed the proto-Galaxy at high redshift. \citet{2023MNRAS.tmp.1847H} showed that a centrally concentrated mass profile is necessary for disk formation with Feedback In Realistic Environments (FIRE) simulation. Metal-poor stars are known to reside in the inner few kiloparsecs of the Milky Way \citep{2013ApJ...767L...9G, 2020MNRAS.491L..11A, 2020MNRAS.496.4964A}, but the current data does not provide a comprehensive picture of this metal-poor "heart" of the Milky Way. However, recent observations taking advantage of the XP spectra from \textit{Gaia} DR3 have revealed an extensive, ancient, and metal-poor population of stars in the inner Galaxy, representing a significant stellar mass \citep{rix_poor_2022}. The early phases of the Milky Way's star formation and enrichment are reflected in the distribution of old and metal-poor stars, which can be a mix of those that formed within the main in situ over-densities of the proto-Galaxy and those that formed in distinct satellite galaxies that later merged with the main body \citep{horta_evidence_2021}. The distinction between in situ formation and accretion can be seen in the abundance patterns of the stars, although at very early epochs, the distinction may become blurry due to the rapid coalescence of comparable mass pieces in major mergers.

Recent observational evidence has shed light on the chemical evolution of the transition period when the disk started forming in the Milky Way. \citet{belokurov_dawn_2022} identified a metal-poor component in the Milky Way called \textit{Aurora} from the APOGEE survey. This component is kinematically hot, with an approximately isotropic velocity ellipsoid and a modest net rotation. They revealed that the in-situ stars in Aurora exhibit a large scatter in elemental abundance ratios and the median tangential velocity of the in-situ stars increases sharply with increasing metallicity when \feh{} is between -1.3 and -0.9. The chemical scatter suddenly drops after this period, signalling the formation of the disk in about one to two Gyr. They proposed that these observed trends in the Milky Way reflect generic processes during the early evolution of progenitors of Milky-Way-sized galaxies, including a period of chaotic pre-disk evolution and subsequent rapid disk formation. Interestingly, many of the most metal-poor {in situ} stars {preceding the disk populations} in their sample have lower \mgfe{} than the traditional high \mgfe{} associated with old stars in the Galaxy (see their figure 6 and 7). 

\citet{conroy_birth_2022} extended the search for in situ halo stars as metal-poor as \feh{} = -2.5 in the H3 survey \citep{2019ApJ...883..107C} and revealed that \alphafe{} gradually declined at low metallicity and rose up around \feh{} $= -1.3$ (see their figure 1). \citep{rix_poor_2022} took advantage of the XP spectra from \textit{Gaia} DR3 and derived reliable metallicity estimates for about two million bright stars, including 18,000 stars with $-2.7 <$ [M/H] $< -1.5$. This massive sample allowed them to present the most comprehensive collection of metal-poor in situ stars in the Milky Way. They showed that the observed \alphafe{}-rise is robust even for stars on near-circular orbits in their sample supplemented by \mgfe{} from APOGEE (their figure 7). Despite using samples from different surveys and selection methods, all of their works showed an unexpected \alphafe{}-rise between \feh{} = -1.3 and -1 where \mgfe{} temporarily drops to an intermediate value between the high- and low-\alphafe{} sequence in the disk. 

The decline in \alphafe{} is expected in all galaxies in time as remnants from intermediate-mass stars ($\sim 3- 8$ \msol{}) explode as SNe Ia and release iron-peak elements unless an increasing amount of massive stars are continually evolving as CCSNe and releasing $\alpha$-elements to balance \alphafe{} due to the rarity of massive stars. However, it is surprising to witness an increase in \alphafe{} after it has started to drop as shown in recent observations. This signals the introduction of a considerable amount of $\alpha$-elements into our Galaxy after SNe Ia have made an impact on the composition of the ISM. The ``simmering'' phase scenario was proposed by \citet{conroy_birth_2022} to explain the rise in \alphafe{}. They fixed the inflow rate constant and adopted a low SFE as \alphafe{} naturally declined due to the onset of SNe Ia to avoid forming too many metal-poor stars. As \alphafe{} reached the lowest point, they increased the SFE in the model by twenty-five times. Many massive stars form and evolve as a result and CCSNe dominate the nucleosynthesis process causing \alphafe{} to rise. However, adjusting the SFE is not the only way to increase short-term star formation and a twenty-five-fold increase is rare in isolated galaxies and requires specific galaxy interactions and mergers \citep{2008A&A...492...31D}. Another feasible scenario is that the \alphafe{}-rise was a symptom of fluctuations in the inflow history. The gas reservoir was kept small as \feh{} rose and \alphafe{} declined and a large amount of fresh gas was brought in through inflow, which achieved a similar effect as changing the SFE. There are additional benefits to this scenario. The SFE could be high throughout the entire proto-Galaxy phase, facilitating the rapid rise in \feh{}. Fuelling star formation with additional fresh gas also reduces the risk of running out of gas, unlike increasing the SFE. The reasoning process is summarized in Figure \ref{fig:flowchat_alpha_fe_rise}, which we will revisit after presenting our results. 

This work aims to investigate the cause behind the \alphafe{}-rise comprehensively with a galactic chemical evolution (GCE) model. GCE models are a computationally efficient approach to studying the evolution of galaxies, particularly their elemental abundances. They use parametric empirical laws to trace the evolution of abundances without directly modelling star formation and gas accretion history as performed in cosmological simulations. They have managed to replicate the age-metallicity and age-\alphafe{} relationship, the stellar density variation in the \fehalphafe{}-plane as {a function of positions in the Milky Way} \citep{2018MNRAS.481.1645M, haywood_revisiting_2019, sharma_chemical_2021, johnson_stellar_2021, chen_chemical_2023}. The remainder of this paper is organized as follows: Section \ref{sec:model} gives an introduction to GCE models and briefly describes the ingredients in our model. Section \ref{sec:results} shows the parameter distribution for the models that satisfy part or all of the descriptions of the observed \alphafe{} behavior. Section \ref{sec:discussion} discusses the implications of our results in light of recent work on the early Milky Way. Section \ref{sec:conclusion} provides a summary of our results.

\section{Model} \label{sec:model}

\begin{table*}
\caption{The values of fixed and free parameters in our GCE model}         
\label{tab:parameters}      
\centering                                      
\begin{tabular}{l l l l}          
\hline\hline                        
Parameter & Meaning & Value \\    
\hline                                   
    $R$ & Radius of the box & 3 kpc  \\
    $N$ & Power in star formation law & 1.4 \\  
    $m_{0, \rm warm}$ & Initial warm gas mass  & 0 \msol{}   \\
    $t_{\rm min, SNeIa}$ & Minimum time delay before first SNe Ia & 150 Myr    \\
    $t_{\rm scale, SNeIa}$ & Timescale for decay of SNe Ia & 1.5 Gyr \\
    $f_{\rm direct}$ & Fraction of supernovae ejecta directly into cold gas & 0.01     \\
    $f_{\rm eject}$ & Fraction of supernovae ejecta lost & 0.2  \\
    $\eta_{\rm SF}$ & Mass-loading factor for gas heated by star formation & 3.0 \\
    $\eta_{\rm SN}$ & Mass-loading factor for gas heated by supernovae & 3.0 \\
    $Z_{\odot}$ & Metallicity of the Sun & 0.0156 \\
    \hline 
    $\epsilon_{\mathrm{SF}}$ & Star formation efficiency constant & $10^{-11 - -9}$ \\
    $m_{0, \rm cold}$ & Initial cold gas mass  & $10^{7.5-9.5}$ \msol{} \\  
    $\dot{m}_{\rm inflow}$ & Inflow rate of fresh gas & $0-5$ \msol{} per year \\
    $t_{\rm cool}$ & Cooling timescale of warm gas & $10^{8-10}$ yr   \\
    $f_{\rm SNIa}$ & Fraction of white dwarfs from stars within (3.2, 8.5) $\rm M_\odot$ that turn into SNe Ia & 0.05-0.2 \\
\hline    
\end{tabular}
\end{table*}

Galactic Chemical Evolution (GCE) models utilize a set of parameters guided by empirical physical laws to simulate the chemical evolutionary trajectory of galaxies. The synthesis of new elements within stars and the subsequent release and recycling of gas {consisting of} these newly produced elements into star formation are critical components of these models. Further mechanisms such as accretion/inflow (introducing fresh gas into the model) and outflow (removing existing ISM) can directly or indirectly shape the chemical evolution depicted by these models. The computational time required to run these models is a fraction of what it takes to trace chemistry in cosmological simulations. Therefore, they allow us to quickly sample an extensive range of parameters to examine the impact of various mechanisms or events on the chemistry of galaxies. 

The model utilized in this work is originally developed by \citet{2017ApJ...835..224A} named \textit{flexCE}. We kept most of the original design but updated a few ingredients. It has many features that make it ideal for exploring the physical conditions of galaxies through elemental abundances. First, it has a multi-phase ISM composed of a cold and warm component, thus relaxing the assumption of instantaneous recycling in most GCE models. The newly synthesized nucleosynthesis yields are not immediately returned to the cold ISM for the next round of star formation. Instead, they are stored in the warm ISM which cooled gradually over time. Second, it has a physical implementation of star formation and evolution. The amount of star formation activity in any given step is determined by the amount of cold ISM at the time and the stars are represented in stellar mass bins with lifetimes. {The SFH in the model is regulated by the mechanisms and thus self-consistent.} The original stellar lifetimes only depended on the progenitor mass through an analytic function. Instead, we sourced stellar lifetimes from PARSEC-1.2S isochrones of various progenitor masses and metallicities \citep{2012MNRAS.427..127B}. The tracking of stellar lifetimes is important for studying the nucleosynthesis inside low-mass stars, including the production of white dwarfs and in turn SNe Ia. 

Third, it uses a complete suite of nucleosynthesis tables. We have updated the model with the most up-to-date tables from \cite{2020ApJ...900..179K} and included magnetorotational supernovae (MRSNe). This allows us to trace up to 83 elements in the model. Lastly, the model has a large selection of free parameters that allow us to fine-tune the strengths of various mechanisms. We can prescribe a function that controls the inflow of fresh gas over time and adjust the mass-loading factor regulating the outflow from star formation and supernovae. The original model assigned the inflowing gas to the cold ISM when it joins the model, but we switched it to the warm ISM so the fresh gas mixes with the existing enriched warm. {The cooling of the inflowing gas is governed by the same cooling timescale for the existing warm ISM.} This change improves the smoothness of chemical evolutionary tracks when a large amount of gas with a different composition joins the model. Section \ref{subsec:setting up} offers a brief description of the model for readers not familiar with the original version. More details can be found in \citet{chen_chemical_2023} where a multi-zone version of this model replicated the variation of the \fehalphafe{} density distribution in various locations of the cross-section plane of the Milky Way. 

\subsection{Setting up our GCE model}
\label{subsec:setting up}
In order to investigate the \alphafe{} behavior at the outset of the Milky Way's evolution, our model is set to run for 1.8 Gyr. This timing is designed to accommodate a subsequent high-\alphafe{} population that could potentially be as ancient as {twelve billion years old \citep{xiang_time-resolved_2022}. We divided the time frame into two phases, the first one lasting one Gyr allowing \alphafe{} to decline and the second one lasting 800 Myr as \alphafe{} rises.} Each time step corresponds to $dt = 30$ Myr, reflecting the lifespan of the longest-living progenitors of CCSNe. Given that the proto-galaxy was likely considerably smaller than the current Milky Way, our circular box is assigned a radius of $R=3$ kpc. {The size of the box is only used for calculating the gas density in our model and does not reflect the physical size of the proto-Galaxy or the distribution of materials within it.} The yield tables from \cite{2020ApJ...900..179K} {linearly} interpolate across the mass bins and then along 1000 grid points in metallicity, ranging from $Z=0$ to $Z=0.06$. The interpolated yields at the nearest metallicity grid point for a stellar mass bin are retrieved when a stellar bin reaches its lifespan, or when a white dwarf explodes. 

We now initialize the components in our GCE model that will house the chemical elements: stars, cold ISM, and warm ISM. The stars are represented by stellar bins ranging from 0.1 to 50 \msol{}, following the initial mass function (IMF) defined below:
\begin{equation} 
\label{eq:imf}
\xi(m) \propto
\begin{cases}
m^{-1.3} & \mathrm{0.1 \leq m < 0.5 \rm M_\odot} \\
m^{-2.3} & \mathrm{m \geq 0.5 \rm M_\odot} \\
\end{cases}
\end{equation}

For stellar bins with masses less than 9 \msol{}, the bin width is 0.1 \msol{}. For those exceeding 9 \msol{}, the bin width expands to 1 \msol{}. These bin sizes correspond to the original progenitor mass grid points in the nucleosynthesis tables. Stars with a mass greater than 9 \msol{} only survive for less than 30 Myr and will {end their lifetime} within a single time step. New stellar bins are created to contain the mass during star formation at each step, and we monitor the remaining mass over time. In addition to mass, we also log the gas composition encapsulated within the stars at the moment of formation. This chemical composition is updated when a stellar bin expires, according to the interpolated yield tables. When a star with a mass between 3.2 and 8.5 \msol{} dies, we reference its remnant mass in the yield tables and add it to a white dwarf reservoir, which is subsequently used to calculate the number of type Ia supernovae (SNe Ia). Both the cold and warm ISM components include 83 entries that correspond to the 83 elements present in our nucleosynthesis tables.

Unlike AGBs and CCSNe which release the yields at the end of their progenitors' lifetime, SNe Ia experience an additional delay time after the formation of white dwarfs (WDs). The total mass of WDs in the galaxy, originating from progenitors with masses between 3.2 and 8.5 \msol{}, divided by the Chandrasekhar limit, determines the maximum number of potential SNe Ia in the model. This number is multiplied by a fraction $f_{\rm SNIa}$ and an exponential delay term $dt / t_{\rm scale, SNeIa}$, where $t_{\rm scale, SNeIa}$ is the delay timescale for SNe Ia. 
\begin{equation} \label{eq:sfr}
    N_{\mathrm{SNIa}, i} = f_{\rm SNIa} \frac{m_{\mathrm{WD}, i}}{1.44 \rm M_\odot} \frac{dt}{t_{\rm scale, SNeIa}}
\end{equation}
The SNe Ia yields utilized in our model are metallicity-dependent and are thus interpolated and applied in a similar fashion to the AGB and CCSNe yields. In summary of the timescales of the three major nucleosynthesis channels, CCSNe explode within thirty Myr (one time step), followed by AGBs from intermediate-mass stars over a few Myr years to a few Gyrs, and lastly succeeded by their WD remnants igniting as SNe Ia.

The SFR in our model is computed using the Kennicutt-Schmidt (KS) law \citep{1959ApJ...129..243S, 1998ARA&A..36..189K, 2012ARA&A..50..531K}, as represented below:
\begin{equation} \label{eq:sfr}
    \frac{d\Sigma_*}{dt} = \epsilon_{\mathrm{SF}} \Sigma_{g}^{1.4} \sim \epsilon_{\mathrm{SF}} \ \left(\frac{m_{i, \mathrm{cold}}}{\pi R^2}\right)^{1.4}
\end{equation}
Here, ${m_{i, \mathrm{cold}}}$ represents the quantity of cold ISM available in the box at time step $i$. The radius of our GCE box is used for calculating the gas density and in turn the SFR in our model. {The materials in the model are assumed to be uniformly distributed. The size of the proto-Galaxy could be rapidly changing but constraining it would require some understanding of the rate of chemical evolution. Here we only aim to capture the average effect of mechanisms. We have set the SFE, $\epsilon_{\mathrm{SF}}$, a free parameter for the star formation mechanism so we decided to fix $R$.} 

\begin{figure*}
\centering
  \includegraphics[width=0.85\linewidth]{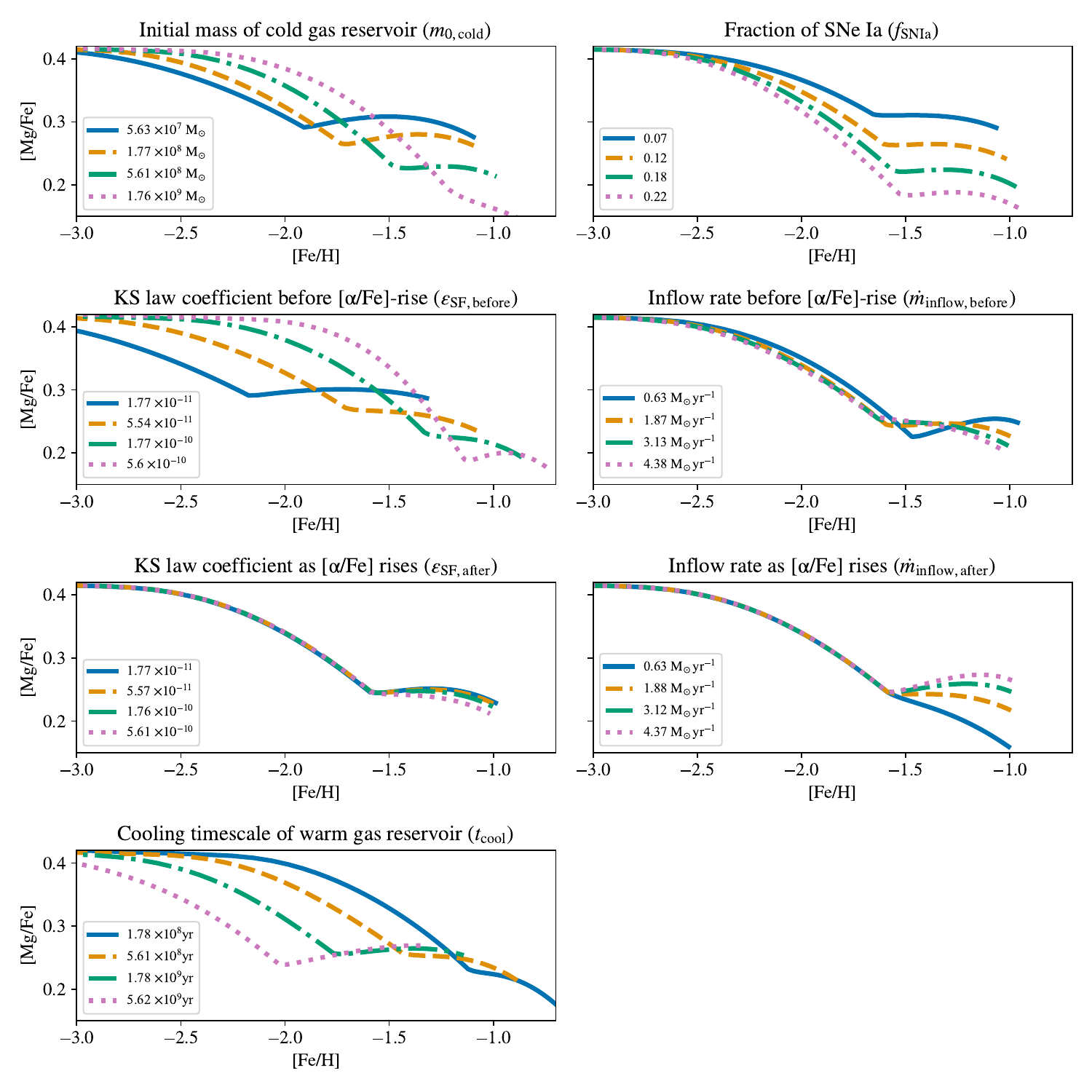}
  \caption{The aggregate effect of the five free parameters on the chemical evolutionary tracks. Each panel represents one free parameter in the GCE model. The SFE ($\epsilon_{\mathrm{SF}}$) and inflow rate ($\dot{m}_{\rm inflow}$) are allowed to change after one Gyr when we expect \alphafe{} to reverse and thus represented in two panels respectively before and after the turning point. Each panel contains four tracks averaged across \mgfe{} within the four quartiles of the parameter range. The median value of each parameter is shown in the legend with the corresponding colour and line style. As the rest of the parameters are drawn randomly in each run, the effect of the other parameters is expected to even out, allowing us to observe the effect of a single parameter. }
  \label{fig:quantile_median_track}
\end{figure*}

The newly formed stellar mass is distributed to stellar mass bins following the IMF mentioned previously, and the corresponding amount of cold ISM is locked inside these stars. Upon reaching their stellar lifetimes and releasing the gas enriched by nucleosynthesis, 1\% is allocated to the cold ISM, 79\% to the warm ISM, and the remaining is assumed to be lost from the model. The warm ISM cools exponentially over a timescale of $t_{\mathrm{cool}}$, during which a fraction equal to $dt/ t_{\mathrm{cool}}$ is transferred to the cold ISM. The cooling of the warm ISM is performed at the beginning of each time step. The process of star formation and evolutionary events will cause a portion of the cold ISM to transition into warm ISM through feedback. We set the mass loading factor $\eta$ at three, which implies that a quantity of cold ISM, equivalent to three times the mass of gas involved in star formation and newly produced yields, will be instantaneously heated {into warm ISM that will eventually be recycled for star formation}. Finally, during each time step, an influx of fresh infalling gas will replenish the warm ISM. 

\subsection{{Exhaustive} parameter exploration}
The conventional approach to creating a GCE model that reproduces specific chemical evolutionary tracks involves manually choosing parameter values and coming up with a standard matching model. However, this method demands strong observational constraints to restrict the resulting GCE scenario, {such as the age-abundance relation and the stellar distribution of specific abundances}. In the case of the Milky Way disk, we have a large amount of observational data to constrain our model. As for the pre-disk Milky Way, our observational data {focus on} a small number of metal-poor stars. We have little knowledge about the star formation history or the overarching properties of the Galaxy {during the first two Gyr after its birth before the formation of the disk}. As a result, we identified five free parameters and ran our model with randomly generated parameter values {until we are able to map the distribution of feasible parameter combinations}. The goal is to explore the physical conditions that may have led to the observed \alphafe{}-rise after \feh{} $\sim -1.3$ and \mgfe{} $\sim 0.2$. The feasible ranges for these parameters are chosen to encompass the values typically utilized for Milky Way studies, as well as those identified during our initial exploration. 

The free parameters are: 
\begin{itemize}
\item  the initial mass of cold interstellar medium ISM, $m_{0, \mathrm{cold}}$; 
\item  the fraction of white dwarfs arising from progenitor stars with initial masses within the range of (3.2, 8.5)\msol{}, eligible for SNe Ia, $f_{\rm SNIa}$; 
\item  the cooling timescale of warm ISM, $t_{\mathrm{cool}}$; 
\item  the SFE, $\epsilon_{\mathrm{SF}}$; and 
\item  the inflow rate, $\dot{m}_{\mathrm{inflow}}$. 
\end{itemize}

{Each of the free parameters plays a crucial role in our GCE model.}

\begin{itemize}
    \item The initial mass of the cold ISM, $m_{0, \mathrm{cold}}$, sets the recorded value of \feh{} after the first round of star formation in the model. We permit it to be between $10^{7.5}$ and $10^{9.5}$ \msol{}.

    \item The fraction of Type Ia supernovae, $f_{\rm SNIa}$, is estimated to be around 5\% \citep{2012MNRAS.426.3282M}, but the actual proportion remains uncertain at high redshift. We permit it to be between 5\% and 20\%. This key parameter influences the evolution of \feh{} and \alphafe{} after several hundred million years when a considerable amount of SNe Ia start producing iron. 

    \item $t_{\mathrm{cool}}$ controls the rate at which newly synthesized metal is returned to the cold ISM for star formation and ranges between $10^8$ to $10^{10}$ years in our model. Determining a cooling timescale for our warm ISM is challenging because it realistically depends on factors such as temperature and metallicity \citep{2012ApJ...759....9K}. However, it should typically range from a few hundred million years to a few billion years. 

    \item $\epsilon_{\mathrm{SF}}$ controls the efficiency of the process through which cold ISM is converted into stars. The SFE constant can be as low as $10^{-11}$ (approximately 1\% per billion years) to cover the possibility of a low SFE scenario. It can also reach $10^{-9}$, comparable to the values estimated by \cite{2008AJ....136.2846B} and \cite{2008AJ....136.2782L} in nearby galaxies.

    \item The inflow rate introduces fresh gas into the model and ranges between 0 and 5 \msol{} per year. The continuous inflow of pristine or metal-poor matter hinders the increase of \feh{} but fuels long-term star formation. 
\end{itemize}

The values of the fixed and free parameters are summarized in Table \ref{tab:parameters}. Parameters $m_{0, \mathrm{cold}}$, $t_{\mathrm{cool}}$, and $\epsilon_{\mathrm{SF}}$ are chosen from a log-uniform distribution, allowing their effects on log-scale elemental abundances to be better observed. The remaining two parameters are selected from a uniform distribution. {The parameter values are drawn independently randomly each time a new run begins without prior memory.} However, replicating the \alphafe{}-rise is not possible if the parameter values remain constant. Once Type Ia supernovae commence, \alphafe{} decreases monotonically, necessitating additional $\alpha$-elements through infall or enhanced star formation to reverse the trend. Consequently, our model allows for {two channels for the production of additional $\alpha$-elements through enhanced star formation activity}. The first channel is through the increase in the SFE. Provided that there is sufficient cold ISM in the model to sustain star formation, the sudden rise in the SFE can lead to a significant amount of CCSNe over a short period of time. Otherwise, the cold ISM is depleted and no further star formation activity is present in the model. The second channel is through gas accretion into the model. {However, if we had kept the inflow gas pristine, the increased inflow would reduce the metallicity of the ISM, which was not observed. We changed its \feh{} ratio to -1.35 after \alphafe{} dropped to the lowest point, which is lower than our target \feh{} $= 1.3$ after one Gyr. } We assumed the inflow gas at this epoch to be $\alpha$-enhanced and assigned its \mgfe{} ratio based on pure CCSNe yields ($\sim 0.41$). The additional fresh gas can also boost the short-term SFR and give rise to CCSNe, even though the SFE does not necessarily rise to facilitate active star formation. Although $\alpha$-enhanced gas could bring in some $\alpha$-elements, the inflow gas is very metal-poor and contain little metal. Thus, the rise in \alphafe{} is primarily driven by star formation. 

\begin{figure}
\centering
  \includegraphics[width=0.8\linewidth]{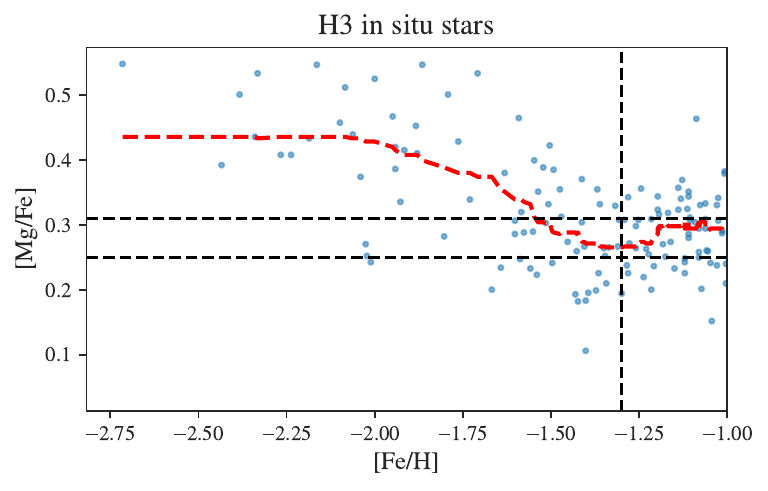}
  \caption{Selected in situ metal-poor stars from the H3 survey in the \feh{}-\mgfe{}-plane. The moving median of \mgfe{} is calculated along \feh{} with a window size of forty and shown in red. The three black dashed lines correspond to three key abundance ratios identified from the median track, \feh{} = -1.3, \mgfe{} = 0.25, \mgfe{} = 0.31.   }
  \label{fig:h3}
\end{figure}

\begin{figure}
\centering
  \includegraphics[width=0.7\linewidth]{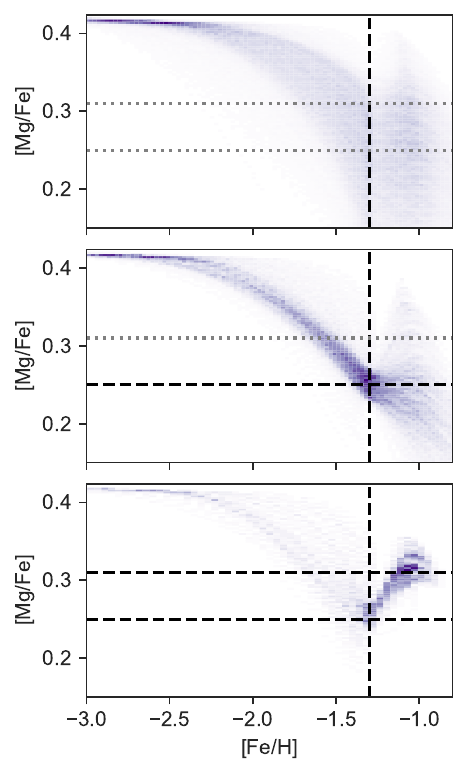}
  \caption{The density distribution of tracks after each selection criterion is incrementally applied to our GCE runs. Tracks in the top panel reach \feh{} of $-1.25 \pm 0.02$ after one Gyr. Tracks in the middle panel reach \mgfe{} of $0.22 \pm 0.02$ after one Gyr in addition to reaching \feh{} of $-1.25 \pm 0.02$ after one Gyr. Tracks in the bottom panel reach \mgfe{} of $0.31 \pm 0.02$ at the last step, besides satisfying the other two criteria. The lines correspond to three key abundance ratios identified from the median \mgfe{}-trend in Figure \ref{fig:h3} used in the selection criteria. In each panel, the applied criterion is shown in dashed lines and the rest in dotted lines.  }
  \label{fig:tracks}
\end{figure}

\section{Results} \label{sec:results}

Before we dig into the \alphafe{}-rise, we will look at each parameter to understand its effect on the chemical tracks. Figure \ref{fig:quantile_median_track} shows the median tracks for the four quartiles of each parameter value within its range. Since $\epsilon_{\mathrm{SF}}$ and $\dot{m}_{\mathrm{inflow}}$ are allowed to change after \alphafe{} reaches the lowest value, there are two panels for each of these two parameters to showcase the tracks before and after the change. {For each parameter in question, we divide its parameter values into four quartiles and obtain a median track over \alphafe{}.} We refer to the point where \alphafe{} reaches the lowest value in the \fehalphafe{}-plane as the \fehalphafe{}-knee, which divides the chemical evolutionary history we study into two phases. Here are the main observations and the rationales behind them from Figure \ref{fig:quantile_median_track}:

\begin{itemize}
\item Increasing the initial mass of cold ISM results in a \fehalphafe{}-knee that is higher in \feh{} and lower in \alphafe{}. Models with more massive initial cold ISM experience have a stronger initial burst of star formation and thus reach a higher metallicity. More active initial star formation also translates to more {early} SNe Ia, causing \alphafe{} to decline {sooner}

\item Increasing $f_{\mathrm{SNIa}}$ results in a \fehalphafe{}-knee that is lower in \alphafe{} and slightly higher in \feh{}. A higher $f_{\mathrm{SNIa}}$ leads to more SNe Ia and more active iron production. The additional iron translates to higher \feh{} and lower \alphafe{}.

\item Increasing $\epsilon_{\mathrm{SF}}$ during the first phase results in a \fehalphafe{}-knee that is higher in \feh{} and lower in \alphafe{}. Models with a higher SFE transform cold ISM into stars and in turn metals more efficiently and thus are more capable of reaching high \feh{}. The more efficient star formation {also leads to more early SNe Ia} and causes more iron to be produced sooner when given the same gas accretion history. 

\item $\epsilon_{\mathrm{SF}}$ during the second phase has no aggregate effect on the \fehalphafe{} tracks. The values of $\epsilon_{\mathrm{SF}}$ before and after the \fehalphafe{}-knee are independent as both are randomly chosen. {$\epsilon_{\mathrm{SF, after}}$ is strongly coupled with other parameters in controlling the star formation mechanism. The SFR does not increase simply because the SFE is turned up. It also depends on whether there is a sufficient amount of gas in the GCE model to sustain the additional star formation activity. The mass of the cold gas reservoir depends on the inflow history before the parameter change and the newly adopted inflow rate after.  }

\item Inflow rate before the \fehalphafe{}-knee has no significant effect on the tracks in the \fehalphafe{}-plane, except when it is very small. {Similar to $\epsilon_{\mathrm{SF, after}}$, $\dot{m}_{\mathrm{inflow, before}}$ is heavily coupled with other parameters.} The elemental abundances reflect the balances of nucleosynthesis channels. Inflow only indirectly affects these channels by influencing the SFH. When the inflow rate is reasonably high, {the SFE or $f_{\mathrm{SNIa}}$ could be low so the star formation activity is suppressed. However, when $\dot{m}_{\mathrm{inflow, before}}$ is very small, the GCE model can be treated as a closed box and thus its chemical enrichment is more effective, evidenced by the higher \feh{} and lower \mgfe{} of the blue track. }

\item Inflow rate after the \fehalphafe{}-knee affects how high \alphafe{} can rise during the second phase. Regardless of the SFH during the first phase, the sudden arrival of fresh gas inevitably causes a large number of massive stars to form and evolve over a short period of time, reversing the declining trend of \alphafe{}. 

\end{itemize}

\begin{figure*}
\centering
  \includegraphics[width=0.9\linewidth]{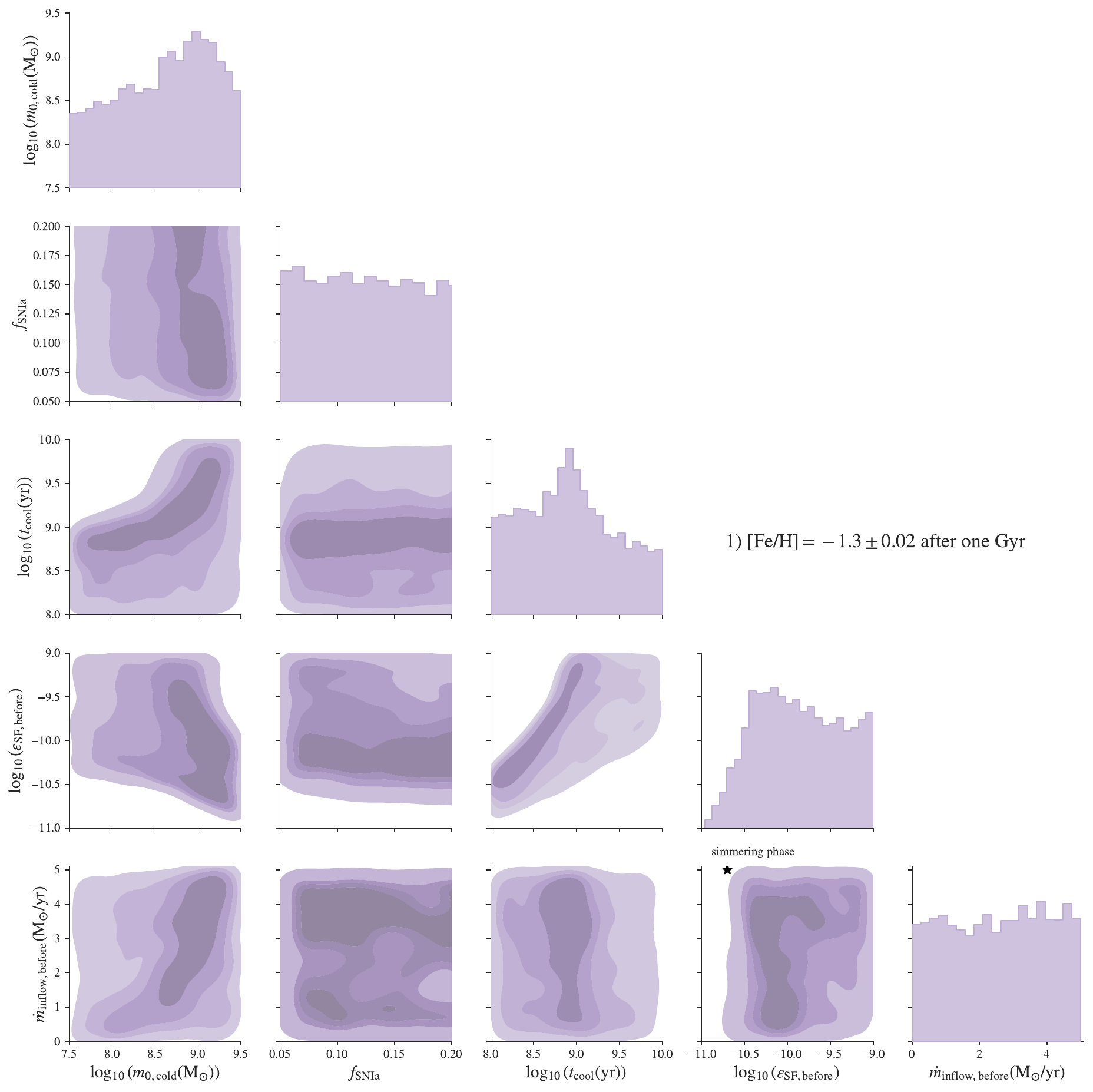}
  \vspace*{-2mm}
  \caption{The distribution of parameter values of the models that reach \feh{} = $-1.25 \pm 0.02$ after one Gyr. The distribution of tracks generated with these parameter values in \feh{}-\mgfe{} is shown in the top panel of Figure \ref{fig:tracks}. The columns and rows correspond to five free parameters in the bottom cell of Table \ref{tab:parameters}, i.e. the initial mass of cold ISM ($m_{0, \mathrm{cold}}$), the fraction of white dwarfs arising from progenitor stars with initial masses within the range of (3.2, 8.5)\msol{} eligible for SNe Ia ($f_{\rm SNIa}$), the cooling timescale of warm ISM ($t_{\mathrm{cool}}$), the SFE constant ($\epsilon_{\mathrm{SF}}$), the inflow rate ($\dot{m}_{\mathrm{inflow}}$). We only show the values of the last two parameters before the \alphafe{}-rise here. The diagonal panels are one-dimensional histograms for each free parameter and the off-diagonal terms are two-dimensional joint distributions between the parameters smoothed by kernel density estimations. The constant inflow rate and the initial SFE of the ``simmering'' phase scenario is marked for reference.}
  \label{fig:pairgrid_1}
\end{figure*}

\begin{figure*}
\centering
  \includegraphics[width=0.9\linewidth]{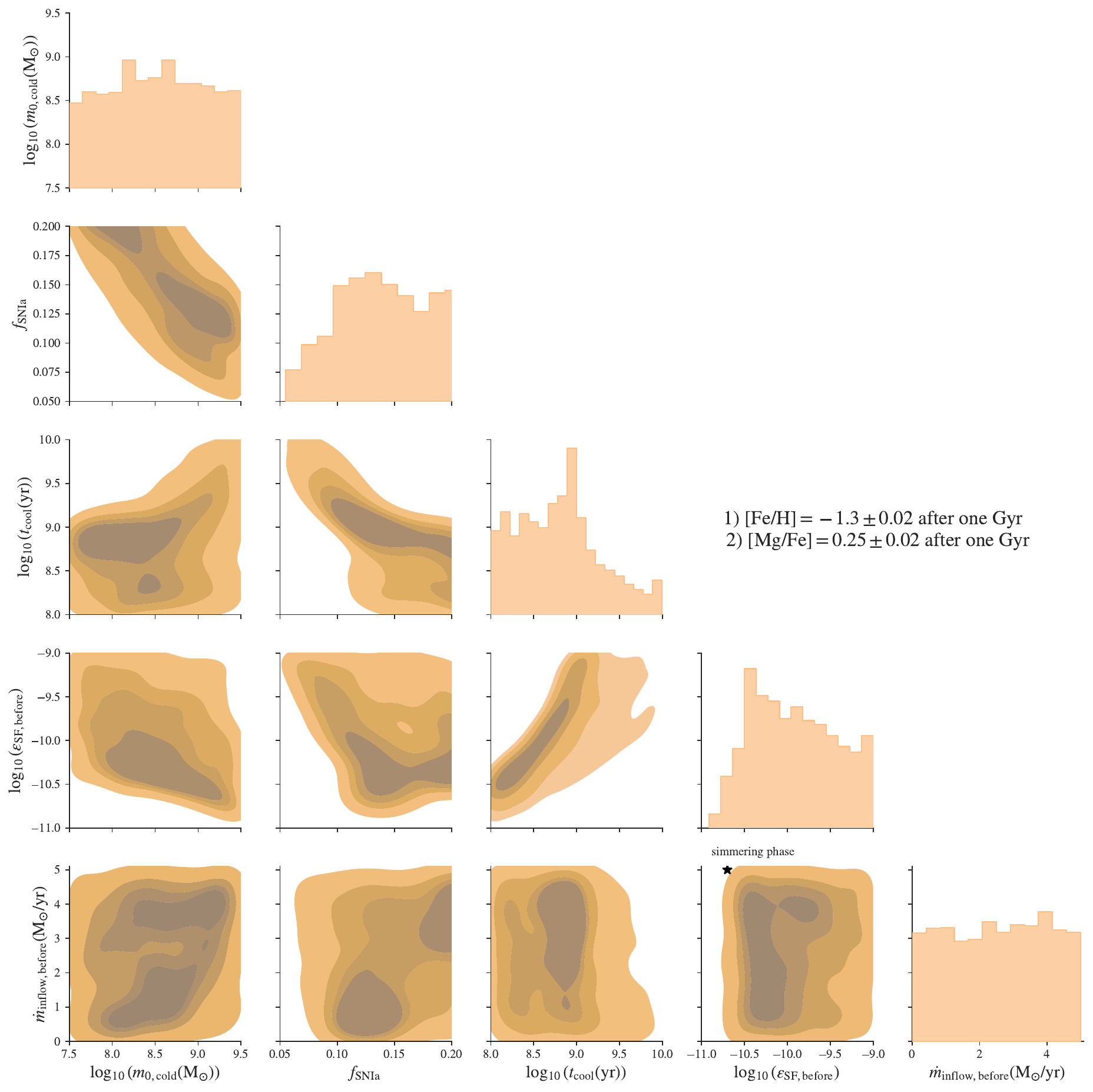}
  \vspace*{-2mm}
  \caption{The distribution of parameter values of the models that reach \feh{} = $-1.25 \pm 0.02$ and \mgfe{} = $= 0.25 \pm 0.02$ after one Gyr in the same style as Figure \ref{fig:pairgrid_1}. The distribution of tracks generated with these parameter values in \feh{}-\mgfe{} is shown in the middle panel of Figure \ref{fig:tracks}.}
  \label{fig:pairgrid_2}
\end{figure*}

\begin{figure*}
\centering
  \includegraphics[width=0.9\linewidth]{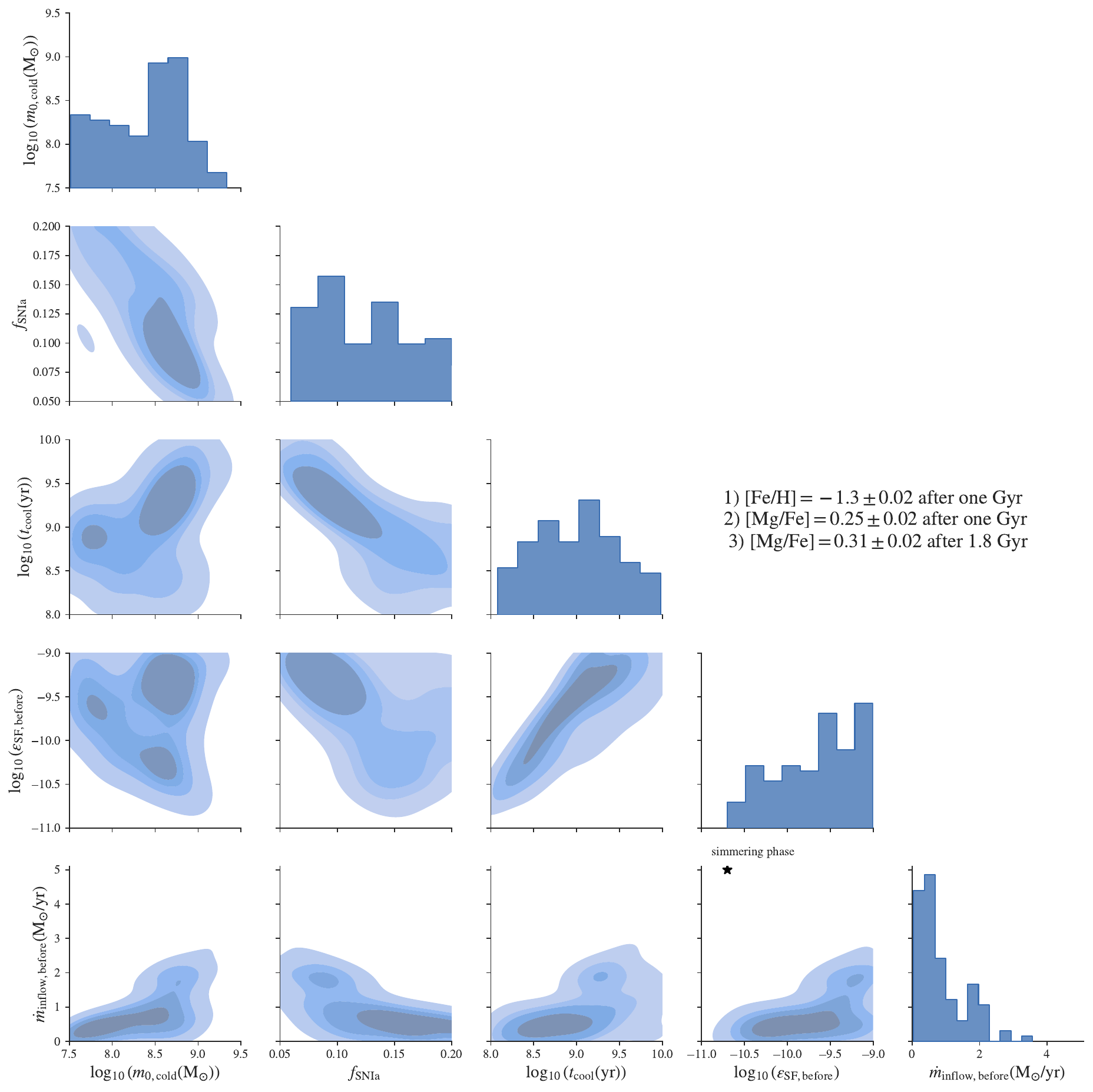}
  \vspace*{-2mm}
  \caption{The distribution of parameter values of the models that reach \feh{} = $-1.25 \pm 0.02$ and \mgfe{} = $= 0.25 \pm 0.02$ after one Gyr as well as reaching \mgfe{} = $0.31 \pm 0.02$ after 1.8 Gyr in the same style as Figure \ref{fig:pairgrid_1} and Figure \ref{fig:pairgrid_2}. The distribution of tracks generated with these parameter values in \feh{}-\mgfe{} is shown in the bottom panel of Figure \ref{fig:tracks}. }
  \label{fig:pairgrid_3}
\end{figure*}

{Although all of the free parameters influence chemical evolution collectively, some parameters can become redundant or crucial, depending on the circumstances.} We identify three key elemental abundances from the H3 in situ sample to characterize the \fehalphafe{}-track we are going to study. Figure \ref{fig:h3} shows the H3 in situ sample in the \feh{}-\mgfe{}-plane. A moving median-\mgfe{} track is fitted to the sample along \feh{} with a window size of forty to reveal the trend in \mgfe{}. A similar trend can be seen in APOGEE data. The three dashed lines correspond to the three key elemental abundances from the \mgfe{}-trend, \feh{} = -1.3, \mgfe{} = 0.25, \mgfe{} = 0.31. We create three selection criteria from these three abundance ratios to isolate models that replicate the rise in \mgfe{}:
\begin{itemize}
    \item \feh{} should reach $-1.3 \pm 0.02$ dex after one Gyr;
    \item \mgfe{} should reach $0.25 \pm 0.02$ dex after one Gyr;
    \item \mgfe{} should reach $0.31 \pm 0.02$ dex at the last time step.
\end{itemize}

These values are arbitrarily chosen by us based on visual checks. The abundances have a large spread in this part of the \fehalphafe{}-plane so it is difficult to pinpoint the key abundance ratios. We leave a small margin for each key ratio because otherwise it {could introduce large variations in the distribution of feasible parameters} and prevent us from drawing physical insights. Nevertheless, the takeaway from this work is the qualitative trends of the parameter values that are immune to small adjustments of these ratios. Figure \ref{fig:tracks} shows the density distribution of tracks in \feh{}-\mgfe{} when each criterion is incrementally applied to all of the tracks generated from within our parameter space. In the top panel, only condition 1) is applied and 6,279 tracks (2.5\% of the total number of runs) remain after the selection. The tracks significantly diverge after \feh{}=-2.25 when iron from SNe Ia starts to influence \mgfe{}. The middle and bottom panels of Figure \ref{fig:tracks} show the distribution of tracks in \feh{}-\mgfe{} when the first two criteria and all three are applied respectively. There are 1,456 tracks (0.58\%) in the middle panel and only 110 (0.044\%) in the bottom. {We will walk through the main results from each criterion that is subsequently applied in the following sections in order.

\subsection{The target \feh{} sets the basic conditions}

Figure \ref{fig:pairgrid_1} displays the distribution of free parameter values {($m_{0, \mathrm{cold}}$, $f_{\rm SNIa}$, $t_{\mathrm{cool}}$, $\epsilon_{\mathrm{SF}}$, $\dot{m}_{\mathrm{inflow}}$)} that satisfy the first criterion (\feh{} = $-1.3 \pm 0.02$ after one Gyr). The diagonal panels showcase one-dimensional histograms of the free parameters, while off-diagonal panels feature {joint distribution} smoothed via kernel density estimations. Examining the diagonal histograms, no preference for $f_{\rm SNIa}$ and $\epsilon_{\mathrm{SF}}$ emerges as their distribution remains uniform, while a preference for $m_{0, \mathrm{cold}} \sim 10^9$ \msol{} and $t_{\mathrm{cool}} \sim 10^{9}$ yr is apparent. As we have seen in Figure \ref{fig:quantile_median_track}, models with a low SFE are unlikely to produce enough iron to meet the desired \feh{} at the \fehalphafe{}-knee as {we have very few models with a SFE less than $10^{-10.5}$}.

Correlations among the remaining free parameter values become evident upon inspecting the {joint distributions}. $m_{0, \mathrm{cold}}$, represented in the first column, determines the amount of star formation and consequently {\feh{} after the first step} in the model. We do not see any significant relationship between $m_{0, \mathrm{cold}}$ and $f_{\rm SNIa}$, but we can see a positive correlation between $m_{0, \mathrm{cold}}$ and $t_{\mathrm{cool}}$ when $m_{0, \mathrm{cold}}$ exceeds $10^{8.7}$\msol{}. Given that we set the power in the Kennicutt-Schmidt law to $1.4$, a larger volume of cold ISM results in a proportionally larger increase in SFR, thereby affecting the amount of metal produced in the {first step and thus the first recorded \feh{} in the model.} The cooling of the warm ISM needs to be extended accordingly to prevent \feh{} from surpassing our target. When $m_{0, \mathrm{cold}}$ falls below $10^{8.7}$\msol{}, the slope is less steep because the metal yield from a low-mass initial cold ISM reservoir is not large enough to require additional cooling; in this case, the \feh{} progression rate can be modulated by other parameters.

Further down the first column, we can see a negative correlation between $m_{0, \mathrm{cold}}$, 
and $\epsilon_{\mathrm{SF}}$. Contrary to the cooling timescale which delays the return of metals into the cold ISM, a high SFE accelerates the conversion of cold ISM into stars and in turn {metals} within the model. When a massive reservoir of cold ISM (high $m_{0, \mathrm{cold}}$) is initially present in the model, the metallicity after one step of star formation is higher and thus the SFE should be restrained to prevent \feh{} from surpassing our specified value. Conversely, with an insignificant initial mass of cold ISM (low $m_{0, \mathrm{cold}}$) and a dependency on infall to accumulate cold ISM, a high SFE is necessary to facilitate the increase in \feh{} over the first one Gyr to reach our \feh{} target in time. Nevertheless, there is a possible low-SFE sequence below $\epsilon_{\mathrm{SF}} = 10^{-10}$ that exhibits a weaker correlation. We will revisit this sequence in the next subsection. 

The bottom panel on the first column shows a significant positive correlation between $m_{0, \mathrm{cold}}$ and $\dot{m}_{\mathrm{inflow}}$. Star formation only converts a few percentages of cold ISM into stars per Gyr, resulting in only a minuscule amount of metal production relative to the amount of inflow gas. Quantitatively, based on the nucleosynthesis tables and the IMF we utilized, every solar mass of core-collapse supernova (CCSN), the primary production site of iron before the onset of SNe Ia, produces about $6.3 \times 10^{-4}$ \msol{} of iron. {The inflow gas during the first one Gyr is assumed to be pristine.} As the cold ISM reservoir is much less massive at this time, even a few \msol{} of pristine gas {per year} can significantly dilute the metal present in the model. Hence, infalling gas primarily inhibits the increase of \feh{} at this time. When $m_{0, \mathrm{cold}}$ is high and the early star formation burst launches \feh{} at a higher value, the inflow rate escalates correspondingly to decelerate the evolution of \feh{} subsequently. 

There is a tight positive correlation between $t_{\mathrm{cool}}$ and $\epsilon_{\mathrm{SF}}$. {In order to reach the same \feh{}, the higher the SFE we adopt for the model, the longer we need to store the nucleosynthesis yields so that the same amount of metals is recycled into the cold ISM}. This relationship only extends to $t_{\mathrm{cool}}$ as high as around one Gyr. When the cooling timescale is longer than one Gyr, the model is forced to adopt a high SFE and regulate the chemical evolution with other parameters. The joint distributions in the remaining panels do not show any significant relationship. Although SNe Ia are typically analogous to iron production when we are studying the chemical evolution in the Milky Way disk, they have a substantial delay time and do not have any significant impact on \feh{} during this phase. However, as we will see soon, \alphafe{} is heavily influenced by SNe Ia. 

In conclusion, our requirement that \feh{} should reach $-1.3 \pm 0.02$ dex in one Gyr selectively favours models featuring a substantial initial cold ISM reservoir ($m_{0, \mathrm{cold}} \approx 10^9$\msol{}), a moderate cooling timescale for the warm ISM ($t_{\mathrm{cool}} \approx 1 \mathrm{Gyr}$), and a relatively elevated SFE ($\epsilon_{\mathrm{SF}} > 10^{-10}$). There is a strong negative correlation between $m_{0, \mathrm{cold}}$ and $\epsilon_{\mathrm{SF}}$ when $m_{0, \mathrm{cold}}$ is higher than $10^{8.5}$ \msol{}. In addition, $m_{0, \mathrm{cold}}$ is positively correlated with $t_{\mathrm{cool}}$ and negatively correlated with $\epsilon_{\mathrm{SF}}$. $\epsilon_{\mathrm{SF}}$ is positively correlated with $t_{\mathrm{cool}}$. The ``simmering'' phase proposed by \citet{conroy_birth_2022} is characterized by a low SFE and a large inflow rate. Their scenario is unlikely based on what we observe in Figure \ref{fig:pairgrid_1} at this stage. Next, we will see constraining \alphafe{} affects the parameter distributions.

\subsection{The frequency of SNe Ia controls the fall of \alphafe{} }

We now explore the parameter space when an additional criterion is imposed on \mgfe{}. Models whose parameter distributions are displayed in green in Figure \ref{fig:pairgrid_2} are required to hit both \feh{} $= -1.3 \pm 0.02$ dex and \mgfe{} $= 0.25 \pm 0.02$ dex at the one Gyr mark. As expected, characteristics identified in Figure \ref{fig:pairgrid_1} reappear in Figure \ref{fig:pairgrid_2}. However, some novel features emerge in the new parameter space, notably in relation to $f_{\rm SNIa}$, the parameter controlling the frequency of SNe Ia. We observe a strong negative correlation between $f_{\rm SNIa}$ and $m_{0, \mathrm{cold}}$, $t_{\mathrm{cool}}$, and $\epsilon_{\mathrm{SF}}$. {Unlike the rest of the free parameters that affect all nucleosynthesis channels, $f_{\rm SNIa}$ only targets SNe Ia.} When $f_{\rm SNIa}$ is low and the synthesized iron is barely sufficient to depress \mgfe{}, {a high $m_{0, \mathrm{cold}}$/$\epsilon_{\mathrm{SF}}$ is required to form as many stars as possible early on so that more SNe Ia explode within our time frame to bring \mgfe{} to our target}. Similarly, a short $t_{\mathrm{cool}}$ ensures that the little iron from SNe Ia is recycled into the cold ISM sooner. Conversely, when $f_{\rm SNIa}$ is high, {the three parameters must work in the opposite direction to prevent too much iron from SNe from lowering \mgfe{} too much}. {The model by \citet{conroy_birth_2022} only has a cold ISM component and effectively has $t_{\mathrm{cool}} = 0$, which is one of the reasons why they consider the ``simmering'' phase feasible.}

The histogram for $f_{\rm SNIa}$ indicates that $f_{\rm SNIa}$ is likely higher than $10\%$, but there is too much of a spread to pin down the exact value, unlike $t_{\mathrm{cool}}$ which shows a strong preference for one Gyr. Meanwhile, we observe shifts in the relationships among the parameters identified earlier. The correlation between $m_{0, \mathrm{cold}}$ and $t_{\mathrm{cool}}$ has become less significant. The high-SFE sequence in the $m_{0, \mathrm{cold}}$-$\epsilon_{\mathrm{SF}}$ plane has been eliminated, leaving behind the low-SFE sequence. A high SFE would form too many stars early on and result in more front-loaded SNe Ia and a faster decline in \alphafe{}. We want to remain in the high-\alphafe{} regime after one Gyr so a moderate SFE is preferred. Regarding the relationship between SFE and $t_{\mathrm{cool}}$, the peak has also gravitated towards a more moderate SFE. During the Milky Way's first Gyr, although the number of SNe Ia is not sufficient to influence \feh{}, it has a substantial effect on \alphafe{}. 

\subsection{The rise of \alphafe{} requires a small existing gas reservoir}

{Finally, we will examine the parameter distributions of the models that not only reached the \fehalphafe{}-knee in the first phase but also managed to raise \alphafe{} during the second phase. Models whose parameter distributions are displayed in red in Figure \ref{fig:pairgrid_3} are required to satisfy all three criteria to complete the \alphafe{}-reversal.} The most distinguishing feature among models achieving a higher \alphafe{} value during the second phase is their inflow rate restricted to less than one \msol{} per year during the first phase. {As low-\alphafe{} gas in a model accumulates through inflow during the first phase, the amount of $\alpha$-elements needed to raise \alphafe{} increases. There are two channels to boost short-term star formation for additional $\alpha$-elements. The first is to accumulate a large gas reservoir and increase the SFE (the ``simmering'' phase). The second is to introduce a massive inflow episode relative to the existing gas reservoir to the model. The contours and histogram in the bottom row of \ref{fig:pairgrid_2} prefer the second scenario. This is not surprising as Figure \ref{fig:pairgrid_1} has indicated from the start that a relatively high SFE is required to meet the \feh{} ratio of the \fehalphafe{}-knee. A large amount of low-\alphafe{} gas present from the first channel makes it extremely difficult to raise \alphafe{}. The relations among the parameters identified in Figure \ref{fig:pairgrid_2} have also been updated. $m_{0, \mathrm{cold}}$ and $t_{\mathrm{cool}}$ become more significantly and strongly correlated. The low-SFE sequence in the $m_{0, \mathrm{cold}}$-$\epsilon_{\mathrm{SF}}$ plane has been replaced by a sequence exhibiting a strong correlation across the entire range. A low $f_{\rm SNIa}$ ($\approx 0.1$) becomes the preferred value. We can observe some additional substructures, highlighting the possibility of sub-scenarios.  }  

\begin{figure*}
\centering
  \includegraphics[width=0.9\linewidth]{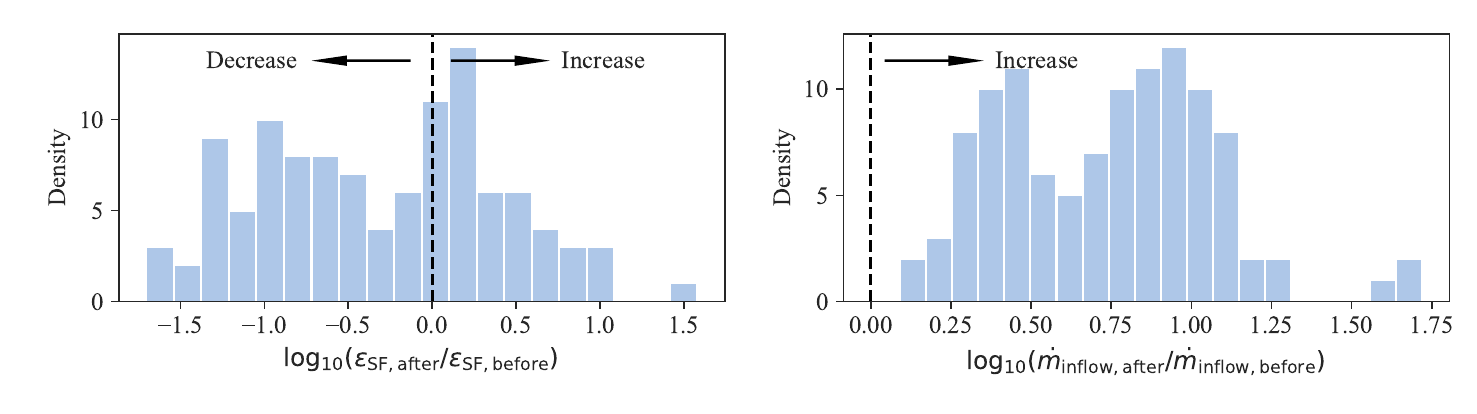}
  \vspace*{-4mm}
  \caption{The change in the parameter values of the SFE and inflow rate before vs. after \alphafe{} reaches the lowest value for models that exhibited a desired rise in \alphafe{}. The dashed line marks where the parameter remains constant. About sixty per cent of the models had a decline in the SFE while all of them had an increase in the inflow rate. }
  \label{fig:change_in_parameters}
\end{figure*}

Examining the change in the SFE and inflow rate reveals that inflow is the primary driver behind the \alphafe{}-rise. Figure \ref{fig:change_in_parameters} illustrates the ratios of the SFEs and inflow rates before and after the \fehalphafe{}-knee for models meeting all three criteria. The ratios are shown in logarithmic scale so zero denotes where the parameter values remain constant. Surprisingly, the SFE declined in about 60\% of the models, as far as 90 \% in some models, even though a higher SFE would facilitate the additional star formation to raise \alphafe{}. This makes inflow much more important in raising the SFR, as all of the models experienced an increase in the inflow rate, as much as forty times in some models. We now have a complete picture of the conditions that caused the \alphafe{}-rise. The proto-Galaxy started with an initial gas reservoir of moderate mass ($10^{8.5} - 10^9$ \msol{}). It maintained a small gas reservoir with little gas inflow ($< 1 $ \msol{} per year) as \feh{} climbed and \alphafe{} declined naturally. However, a massive amount of inflow joined the proto-Galaxy around \feh{} $= -1.3$, causing \alphafe{} to rise as a result of the enhanced star formation activity. Through the entire proto-galaxy epoch, the SFE is high ($> 10^{-10}$) and the frequency of SNe Ia {could be higher} than that measured from the local group which is $3 - 10$ \% \citep{2012MNRAS.426.3282M, 2012PASA...29..447M}. The recycling of metal is relatively efficient ($t_{\mathrm{cool}} \approx $ one Gyr). 

\section{Discussion} \label{sec:discussion}

{The purpose of this work is to identify the parameter combinations that will cause \alphafe{} to rise in the \fehalphafe{}-plane from a galactic chemical evolution model and infer the physical condition of the Milky Way before the formation of the disk. We are primarily dealing with two major nucleosynthesis channels in the \fehalphafe{}-plane: CCSNe and SNe Ia. Since only CCSNe synthesize $\alpha$-elements, the most logical explanation behind the rise in \alphafe{} is a boost in the SFR. The question then is what causes the increase in the SFR. 

There are two channels to boost star formation in the model, which correspond to two scenarios of what happened when the disk formed in the Milky Way. The first channel is to increase the SFE. However, a higher SFE would not translate to a high SFR unless there is a substantial amount of ISM to sustain star formation. Thus, the first scenario entails a high inflow rate of fresh gas to build up a large gas reservoir before disk formation and an elevated SFE as soon as the disk forms. The second channel is to introduce additional inflow, which is metal-poor and $\alpha$-enhanced at this time, to expand the gas reservoir and make available more gas for star formation. However, this requires the existing reservoir to be limited in mass, or the new inflow rate becomes unrealistically large. The ingredients for these two channels are organized in the flowchart in Figure \ref{fig:flowchat_alpha_fe_rise}. We performed an experiment by running our model 250,000 times with randomly generated parameters, two of which, the SFE and inflow rate, were allowed to change after one Gyr of runtime was reached. The results preferred the second scenario under which the inflow was initially suppressed. 

We can now clarify the outcomes of our flowchart in Figure \ref{fig:flowchat_alpha_fe_rise} and identify the most likely scenario for the \alphafe{}-rise. As we mentioned, the rise in \alphafe{} is achieved by a corresponding rise in the SFR, which requires at least one of two parameters, the SFE and inflow rate, to rise. It is impossible to reverse the declining trend of \alphafe{} if neither parameter increases. According to the narrative of the ``simmering'' phase, the inflow rate remains high during the entire proto-Galaxy phase while the enhanced SFE boosts the SFR. This scenario was deemed feasible due to the design of the model by \citet{2020MNRAS.498.1364J}. Their model uses a return function to determine the amount of evolved stellar mass instead of tracking stellar lifetime and does not have a multi-phase ISM. These ingredients significantly expedite the production and recycling of metals. \citet{conroy_birth_2022} doubled the yields of SNe Ia, ran the model for 3.7 Gyr (twice as long as ours), and boosted the SFE by twenty-five times to replicate the \alphafe{}-rise. Our results show that a large gas reservoir makes it difficult for \feh{} to reach $-1.3$ in one Gyr and difficult for \alphafe{} to rise, even though the SFE rose as much as ten times within our predefined range in some runs of our model. Thus, the amount of gas in the gas reservoir must have remained low while \alphafe{} was declining. Since there is no existing gas to sustain enhanced star formation, the inflow of fresh gas becomes a necessary condition, while the SFE can rise or fall, as long as it is above a certain threshold. 

\subsection{Further constraining parameters of proto Milky Way}

{We adopted a limited time frame (1.8 Gyr) and stringent criteria to replicate the \alphafe{}-rise in our GCE model in this work, i.e. \feh{} $ = -1.3 \pm 0.02$, \mgfe{} $ = 0.25 \pm 0.02$ after one Gyr and \mgfe{} $ = 0.31 \pm 0.02$ after 1.8 Gyr. {The time frame was chosen to accommodate the possibility of a twelve-Gyr-old thick disk. The three elemental ratios were chosen to} constrain parameter confounding in order to more clearly identify the parameter combinations that would produce the \alphafe{} trend we approximated from observations. Changing the key abundance ratios in the criteria or allowing larger margins could shift the parameter distributions, but it will not invalidate the qualitative trends we observed in Figure \ref{fig:pairgrid_1} and \ref{fig:pairgrid_2}. However, as the precision of the abundance measurements improves or the age estimates for the stars in question become available, we expect to come up with a quantitative approximation of the track rather than three abundance ratios. We can then further fine-tune our parameter choices by controlling the accumulation rate of metal in the model. Nevertheless, there are a few ways to further refine our parameters with tools other than GCE models. The contours in Figure \ref{fig:pairgrid_2} reveal strong correlations among the free parameters. If we can pin down the cooling timescale of the warm ISM in the early universe, we could retrieve the rest of the ``free'' parameters from the correlations. Except for the inflow rate, the remaining four parameters {are well constrained} once one of them is defined to replicate the \alphafe{}-rise.}

{There are other properties we can examine, besides the chemical evolutionary track. Although models with different SFEs can achieve the same key abundance ratios, their stellar chemical distribution and global properties are vastly different.} Figure \ref{fig:abundance_dist} shows the Metallicity Distribution Function (MDF) and the \alphafe{} Distribution Function (ADF) of the models satisfying all three criteria for the \alphafe{}-rise, with a colour coding corresponding to their SFE. {We chose SFE for the colour coding because it is measurable from observations or simulations.} Models with lower SFEs conspicuously contain fewer metal-poor high-\alphafe{} stars, as the SFR is much lower in the initial stage immediately after the birth of the galaxy. Conversely, models boasting exceptionally high SFEs display a dual-peaked distribution. {The cold ISM reservoirs in these models are exhausted by the initial star formation burst and require the replenishment of inflow before star formation can resume. The exact ratio between the two peaks and the width of the gap between them are determined by the parameters we assign to the model.} Figure \ref{fig:gas_stellar_mass} shows the evolution of the global properties of the same models in Figure \ref{fig:abundance_dist} over time, colour-coded again by SFE. {Specifically, we show the mass of cold ISM, warm ISM, and stars as well as the mass ratio between stars and cold gas over running time, and the star formation history.} Except for the stellar mass panel, the rest of the panels show strong correlations of the properties with the SFEs. As observation evidence becomes more abundant in the future {or procedures are developed to reduce the effect of the selection function}, it is possible for us to constrain the SFE more precisely based on the chemical distribution or the global property of the in-situ population in the early Milky Way. We will subsequently discuss the complexities of our findings in the context of prior relevant studies.

\subsection{The inflow that fueled the Milky Way}

There are two physically distinct regimes of gas accretion onto galaxies: a ``cold'' filamentary mode in which warm gas ($< 10^{5}$ K) accretes along filaments that can penetrate to the inner regions of halos and a ``hot'' flow mode wherein hot diffuse gas in extended halos cools over time \citep{2005MNRAS.363....2K, 2009MNRAS.396.2332K}. {The cold accretion mode is more common in smaller galaxies and the hot mode is more common in massive galaxies.} The important role of gas accretion has been established by cosmological simulations but the properties of accreted gas have been difficult to quantify because the baryon cycle is a complex process that involves the continuous interplay of gas inflow, outflow, and star formation \citep{2009Natur.457..451D, 2011MNRAS.414.2458V, 2014MNRAS.440..920L, 2015MNRAS.448...59N, 2018MNRAS.478..255C, 2020MNRAS.494.3971M, 2020MNRAS.498.1668W}. The prototypes of Milky Way analogs in cosmological simulations at high redshift experience rapid gas accretion through cold streams with large angular momentum which is subsequently transferred to the existing halo, contributing to the increased spin of these galaxies \citep{2011ApJ...738...39S, 2012MNRAS.423.1544S, 2015MNRAS.449.2087D}. However, before the filaments with different directions become aligned and settle into disks, Milky Way prototypes take the shape of spheroids characterized by an extended profile and violent kinematics \citep{2012MNRAS.423..344R, 2013ApJ...769...74S, 2013ApJ...763...26O, 2019MNRAS.487.4424O, 2013ApJ...773...43B, 2021MNRAS.503.1815B, 2019MNRAS.486.1574M}. As these galaxies accrete more gas, they become massive enough to support a hot halo which subsequently triggers the transition from ``cold'' mode accretion to ``hot'' mode or cooling mode accretion, accompanied by the formation of a gaseous disk \citep{2020MNRAS.493.4126D, 2021ApJ...911...88S, 2022MNRAS.514.5056H, 2023MNRAS.519.2598G}. 

Our work is closely connected to a series of recent works on the pre-disk Milky Way. \citet{belokurov_dawn_2022} first identified a large number of stars before the coherent disk in the Galaxy formed and named this population \textit{Aurora}. The features of this population are consistent with the scenario under which stars form in cold filaments that are rapidly accreted onto the Galaxy. The stars have a large scatter in elemental abundances which could be caused by the diverse conditions under which nucleosynthesis took place. Its spheroidal spatial distribution and isotropic velocity ellipsoid are expected from simulated galaxies that went through the chaotic phase of evolution. Additionally, these stars showed a strong positive correlation between metallicity and tangential velocity, which is a signature of the filaments transferring their high angular momentum to our Galaxy. \citep{conroy_birth_2022} extended the metallicity coverage of this population, especially towards the metal-poor end, and presented us a more complete picture of the \alphafe{}-rise. The dynamical aspect of these works is corroborated by \citet{2023MNRAS.523.6220Y} that showed that the orbits of in situ stars are closely related to their respective formation epoch from simulated Milky Way-mass galaxies in FIRE. {Our accretion scenario suggests that the cold accretion should take time to ramp up in the proto-Galaxy. }

Our results offer additional evidence from the perspective of chemical evolution that the brief \alphafe{}-rise in the \fehalphafe{}-plane is a signature of massive inflow that ended the prototype phase of the Milky Way and initiated the physical process through which the disk later formed. Although our GCE model does not simulate the dynamical features of our Galaxy during its earliest epoch, it is flexible with a wide selection of parameters governing the chemical evolution to explore the conditions behind the \alphafe{}-rise. Unlike traditional GCE studies that present one model that demonstrates the most likely scenario, we ran our model 250,000 times to generate the distribution of parameter values. Although it is reasonable to expect the SFE to rise to facilitate enhanced star formation, our model showed that inflow must be suppressed until the rise of \alphafe{}, while the SFE can rise or fall. This scenario agrees with the cold mode accretion that fueled the formation of the early Milky Way suggested by \citet{belokurov_dawn_2022}. \citet{rix_poor_2022} estimated their in situ metal-poor sample to have a stellar mass $\mathrm{M_*} > 10^{8}$ \msol{} which also agrees with the median stellar mass of the proto-Milky Way from our model (see Figure \ref{fig:gas_stellar_mass}). 

\begin{figure}
\centering
  \includegraphics[width=0.7\linewidth]{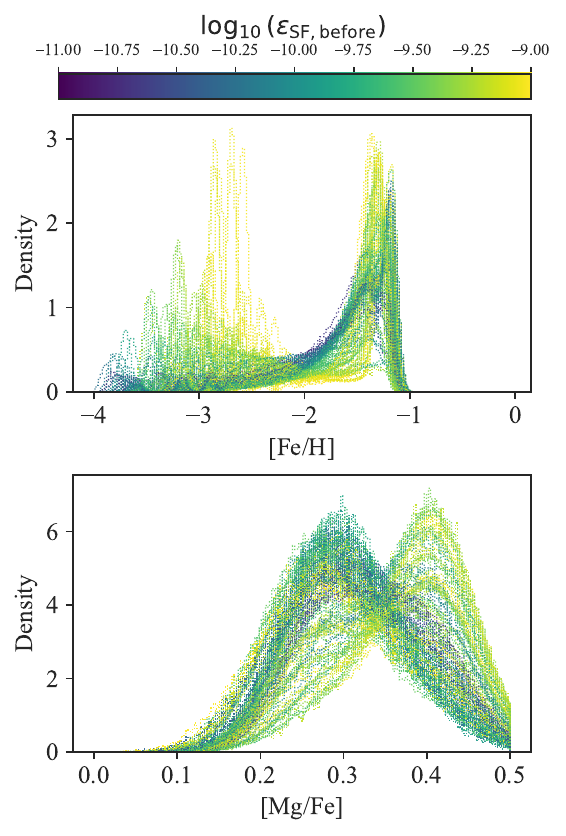}
  \caption{Stellar density distribution of [Fe/H] and [Mg/Fe] in models with an \alphafe{}-rise at the last time step (1.8 Gyr of runtime) colour-coded by the SFE. Models with different SFEs have distinct chemical distributions. }
  \label{fig:abundance_dist}
\end{figure}

\begin{figure}
\centering
  \includegraphics[width=0.7\linewidth]{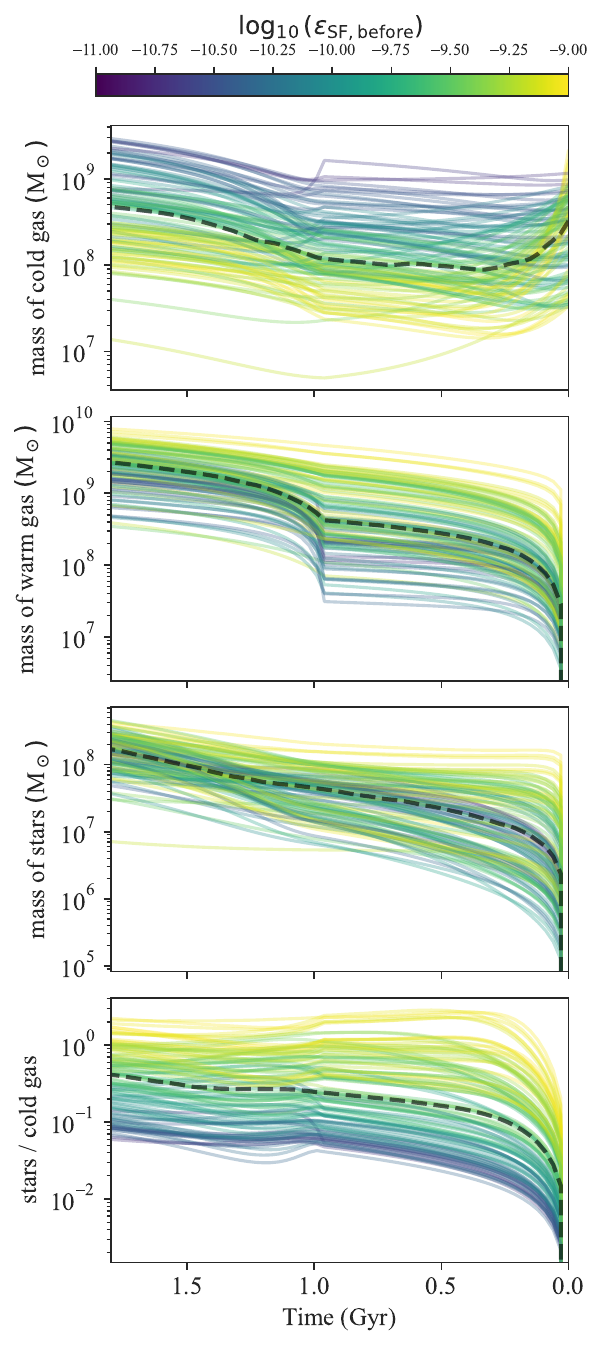}
  \caption{{The masses of cold gas, warm gas, stellar mass, the mass ratios between stars and cold gas over the course of the GCE runs, and the star formation histories for the models exhibiting an early \alphafe{}-rise.} The panels from top to bottom show the evolution of the total cold gas mass, warm gas mass, stellar mass, and the ratio between cold gas and stellar mass. }
  \label{fig:gas_stellar_mass}
\end{figure}

\subsection{Contamination from accreted stellar structures}

It is possible that the stellar sample exhibiting the observed \alphafe{}-rise is contaminated by accreted stellar structures. Distinguishing in situ and accreted stars in the halo of the Milky Way is primarily done through the analysis of their kinematic and chemical properties, but it is unclear to what extent the properties are related to the birth origin of stars. The kinematic properties of stars, such as their orbits and angular momentum characteristics, can provide insights into their origin and formation history. Accreted stars in the halo are typically associated with tidal debris from disrupted satellite galaxies, such as Sagittarius and Gaia-Sausage-Enceladus, and have highly inclined or eccentric orbits \citep{belokurov_co-formation_2018, 2018_helmi_enceladus}. The presence of disrupted star clusters in the stellar halo is also indicative of an accreted component \citep{2018MNRAS.481.3442M, 2018ApJ...862..114S, 2021ApJ...909L..26B}. Nevertheless, even among in situ stars in simulated Milky Way analogs, there is a wide range of dynamical features \citep{2023MNRAS.523.6220Y}. Elemental abundances can also be used to distinguish in situ and accreted stars in the halo. The distinct abundance patterns of accreted stars in the \fehalphafe{} or \mgmnalmn{} plane suggest an accreted origin, indicating that their birth material was enriched in a lower mass potential well, such as a satellite galaxy \citep{2015MNRAS.453..758H, 2015ApJ...802...48L, belokurov_biggest_2019, feuillet_selecting_2021}). The question of distinguishing the birth origin of stars is probably moot during the proto-Galaxy phase. \cite{2018MNRAS.480..652E} studied the distribution of ancient stars in simulated Milky Way analogs in detail and found that most of the oldest stars are accreted through hierarchical assembly. At $z = 5$ ($\sim 1.1$ Gyr after the Big Bang), the main progenitors of the Milky Way analogs contained only half of the old stars in stellar mass or less. If the observed \alphafe{}-rise were manifested in one or multiple accreted structures, they likely represented a significant portion of the stellar mass of the proto-Galaxy.  

\subsection{Implications on future GCE studies of the Milky Way disk}

One of the challenges to studying the Milky Way disk with GCE models is setting the initial conditions when the disk first formed. The standard approach consists of initializing a GCE model with a reservoir of pristine gas and growing the reservoir over time with inflow which is also pristine or very metal-poor (\feh{} < -1). Two issues arise from this approach. The first is that the time it takes for \feh{} to reach a relatively high value, around $-0.5$ for the thick disk, is often longer than the age of the thick disk stars, around ten to twelve Gyr. Only 2.5\% of our runs reached \feh{} $\approx -1.3$ in one Gyr and fewer reached \feh{} $\approx -1.3$ at 1.8 Gyr. The SFE and other parameters can be adjusted so that \feh{} rises faster. However, a substantial amount of gas needs to be added to the reservoir via inflow early to achieve a high SFR during the formation of the thick disk, which is a common ingredient needed to replicate the \alphafe{}-bimodality in the disk \citep{2015A&A...580A.126K, minchev_chemodynamical_2013, johnson_stellar_2021, chen_chemical_2023}. The combination of a high SFE and a massive gas reservoir leads to too many metal-poor stars in the model. The second issue is how \alphafe{} managed to remain high for an extended period of time until the thin disk started forming. The \alphafe{} value of the thick disk ranges between 0.2 and 0.4. We chose \mgfe{} in this work which has an \alphafe{}-ceiling of around 0.41 according to the yields of CCSNe. The \alphafe{} value in a large fraction of runs in the top panel of Figure \ref{fig:tracks} have gone down by at least 0.1 dex after one Gyr when the thick disk had started to form according to \citet{xiang_time-resolved_2022}. As more SNe Ia explode, \alphafe{} would decrease even further and deviate from the ratio associated with the high-\alphafe{} population. 

The scenario for the proto-Galaxy outlined in our work solves the above issues and allows future GCE studies of the Milky Way disk to set initial conditions in a self-consistent approach. The inflow is initially suppressed in the model, helping metals accumulate without being diluted by metal-poor gas. We had a fixed time stamp for \feh{} to reach around -1.3 at one Gyr. However, the model is capable of reaching a higher metallicity within the same amount of time in Figure \ref{fig:quantile_median_track}. In the middle panel of Figure \ref{fig:gas_stellar_mass}, most of our models formed $10^8$ \msol{} of stars, less than 1 \% of the total stellar mass of our Galaxy. The small gas reservoir facilitates the rapid rise of \feh{} while keeping the SFR low. As for the second issue, \alphafe{} inevitably drops below the desired high value, but we rejuvenate the gas reservoir with a higher inflow rate. Since the initial gas reservoir is limited in mass, the large amount of fresh gas leads to a new episode of star formation, raising \alphafe{} and further elevating \feh{}. This scenario allows us to achieve a high \alphafe{} value in our model even after two to three Gyr to correctly reproduce the elemental abundances of the thick disk. 

\section{Conclusion} \label{sec:conclusion}

In this study, we deepened our understanding of the [$\alpha/\text{Fe}$]-rise observed in H3 and APOGEE data through a comprehensive investigation using our GCE Model. The main findings are as follows:

\begin{itemize}

\item The [$\alpha/\text{Fe}$]-rise is principally driven by gas inflow, thus adding a chemical evolution perspective to theories surrounding the Milky Way disk's spin-up phase. Specifically, the ISM of the proto-Galaxy initially appeared isolated to facilitate a quick rise in [Fe/H] and later underwent rapid gas accretion.

\item Contrary to prior studies, our results show that the SFE does not play a deterministic role in the [$\alpha/\text{Fe}$]-rise, even though in theory the rise in the SFE should facilitate the [$\alpha/\text{Fe}$]-rise. Interestingly, SFE could either increase or decrease, yet still result in the observed [$\alpha/\text{Fe}$] rise under certain conditions.

\item The models suggest that the earliest proto-Galaxy had an initial gas reservoir ranging from $10^{8.5}$ to $10^9 \text{M}_\odot$, along with efficient cooling of the warm ISM on a timescale of one Gyr.

\item Our model also indicates a marginally higher frequency of SNe Ia than currently observed. This involves at least $5-10\%$ of white dwarfs originating from stars within [3, 8] $\text{M}_\odot$, implying a higher binary fraction.

\end{itemize}

Our model provides a coherent framework that addresses several key questions in the chemical evolution of the Milky Way

\begin{itemize}

\item Our model addresses the lack of metal-poor stars by initially suppressing gas inflow, allowing for metal accumulation without dilution from metal-poor gas. The small initial gas reservoir and low SFR (not necessarily low SFE) also mean that fewer metal-poor stars are formed, which aligns well with observations.

\item Our model accommodates the observed plateau in [$\alpha/\text{Fe}$] values by introducing a higher inflow rate after the initial phase. This reinvigorates star formation and allows for a sustained high [$\alpha/\text{Fe}$] level, yet lower than the original plateau. This explains why the thick disk's [$\alpha/\text{Fe}$] values plateau at around 0.3 dex, a level lower than those predicted by CCSNe ($\sim$ 0.4 dex)

\item By allowing for a rejuvenated gas reservoir with higher inflow rates, our model captures the startup of the disk formation from the perspective of chemical evolution and collaborates proposed scenarios for the formation of the thick disk.
\end{itemize}

Finally, we note that the [$\alpha/\text{Fe}$]-rise alone leaves considerable ambiguity in model parameters, leading to some degree of degeneracy. Our findings suggest that future observations—whether focused on gas mass or the distribution function in [Fe/H] and [$\alpha/\text{Fe}$]—could significantly improve our understanding of the ancient history of our Galaxy. 

In a broader context, the methodology and insights gleaned from this study extend beyond the Milky Way. Similar approaches could be applied to other intriguing galaxies, such as M31 and the Large Magellanic Cloud, the latter of which have also shown fascinating chemical evolution patterns \citep{2020ApJ...895...88N}. This opens up a wide field for future research. Our work emphasizes the ongoing importance of studying the chemical evolution of galaxies as a vital tool for understanding not only the Milky Way but also the Local Group at large. This significance is poised to grow as new generations of spectroscopic surveys come online and as forthcoming 30-m class telescopes continue to expand our observational horizons.

\section*{Acknowledgements}

YST acknowledges financial support from the Australian Research Council through DECRA Fellowship DE220101520. MRH acknowledges financial support from the Australian Research Council through the Laureate Fellowship awarded to Prof. Joss Bland-Hawthorn.

\section*{Data Availability}

The inclusion of a Data Availability Statement is a requirement for articles published in MNRAS. Data Availability Statements provide a standardised format for readers to understand the availability of data underlying the research results described in the article. The statement may refer to original data generated in the course of the study or to third-party data analysed in the article. The statement should describe and provide means of access, where possible, by linking to the data or providing the required accession numbers for the relevant databases or DOIs.



\bibliographystyle{mnras}
\bibliography{chemical_evolution} 

\bsp	
\label{lastpage}
\end{document}